\documentclass[%
 reprint,
superscriptaddress,
 amsmath,amssymb,
 aps,
]{revtex4-2}

\usepackage{graphicx}
\usepackage{dcolumn}
\usepackage{bm}
\usepackage{upgreek}
\usepackage{soul,color,xcolor}
\usepackage{ulem}

\begin{document}

\preprint{APS/123-QED}

\title{Singlet-triplet-state readout in silicon metal-oxide-semiconductor double quantum dots}

\author{Rong-Long Ma}
\affiliation{CAS Key Laboratory of Quantum Information, University of Science and Technology of China, Hefei, Anhui 230026, China}%
\affiliation{CAS Center for Excellence in Quantum Information and Quantum Physics, University of Science and Technology of China, Hefei, Anhui 230026, China}
\author{Sheng-Kai Zhu}
\affiliation{CAS Key Laboratory of Quantum Information, University of Science and Technology of China, Hefei, Anhui 230026, China}%
\affiliation{CAS Center for Excellence in Quantum Information and Quantum Physics, University of Science and Technology of China, Hefei, Anhui 230026, China}
\author{Zhen-Zhen Kong}
\affiliation{Integrated Circuit Advanced Process R$\&$D Center, Institute of Microelectronics, Chinese Academy of Sciences, Beijing 100029, P. R. China}
\author{Tai-Ping Sun}
\affiliation{CAS Key Laboratory of Quantum Information, University of Science and Technology of China, Hefei, Anhui 230026, China}%
\affiliation{CAS Center for Excellence in Quantum Information and Quantum Physics, University of Science and Technology of China, Hefei, Anhui 230026, China}
\author{Ming Ni}
\affiliation{CAS Key Laboratory of Quantum Information, University of Science and Technology of China, Hefei, Anhui 230026, China}%
\affiliation{CAS Center for Excellence in Quantum Information and Quantum Physics, University of Science and Technology of China, Hefei, Anhui 230026, China}
\author{Yu-Chen Zhou}
\affiliation{CAS Key Laboratory of Quantum Information, University of Science and Technology of China, Hefei, Anhui 230026, China}%
\affiliation{CAS Center for Excellence in Quantum Information and Quantum Physics, University of Science and Technology of China, Hefei, Anhui 230026, China}
\author{Yuan Zhou}
\affiliation{CAS Key Laboratory of Quantum Information, University of Science and Technology of China, Hefei, Anhui 230026, China}%
\affiliation{CAS Center for Excellence in Quantum Information and Quantum Physics, University of Science and Technology of China, Hefei, Anhui 230026, China}
\author{Gang Luo}
\affiliation{CAS Key Laboratory of Quantum Information, University of Science and Technology of China, Hefei, Anhui 230026, China}%
\affiliation{CAS Center for Excellence in Quantum Information and Quantum Physics, University of Science and Technology of China, Hefei, Anhui 230026, China}
\author{Gang Cao}
\affiliation{CAS Key Laboratory of Quantum Information, University of Science and Technology of China, Hefei, Anhui 230026, China}%
\affiliation{CAS Center for Excellence in Quantum Information and Quantum Physics, University of Science and Technology of China, Hefei, Anhui 230026, China}
\affiliation{Hefei National Laboratory, University of Science and Technology of China, Hefei 230088, China}
\author{Gui-Lei Wang}
 \email{guilei.wang@bjsamt.org.cn}
 \affiliation{Integrated Circuit Advanced Process R$\&$D Center, Institute of Microelectronics, Chinese Academy of Sciences, Beijing 100029, P. R. China}
 \affiliation{Hefei National Laboratory, University of Science and Technology of China, Hefei 230088, China}
  \affiliation{Beijing Superstring Academy of Memory Technology, Beijing 100176, China}
\author{Hai-Ou Li}
 \email{haiouli@ustc.edu.cn}
 \affiliation{CAS Key Laboratory of Quantum Information, University of Science and Technology of China, Hefei, Anhui 230026, China}%
\affiliation{CAS Center for Excellence in Quantum Information and Quantum Physics, University of Science and Technology of China, Hefei, Anhui 230026, China}
\affiliation{Hefei National Laboratory, University of Science and Technology of China, Hefei 230088, China}
\author{Guo-Ping Guo}
\affiliation{CAS Key Laboratory of Quantum Information, University of Science and Technology of China, Hefei, Anhui 230026, China}%
\affiliation{CAS Center for Excellence in Quantum Information and Quantum Physics, University of Science and Technology of China, Hefei, Anhui 230026, China}
\affiliation{Hefei National Laboratory, University of Science and Technology of China, Hefei 230088, China}
\affiliation{Origin Quantum Computing Company Limited, Hefei, Anhui 230026, China}

\date{\today}

\begin{abstract}
High-fidelity singlet-triplet state readout is essential for large-scale quantum computing. However, the widely used threshold method of comparing a mean value with the fixed threshold will limit the judgment accuracy, especially for the relaxed triplet state, under the restriction of relaxation time and signal-to-noise ratio. Here, we achieve an enhanced latching readout based on Pauli spin blockade in a Si-MOS double quantum dot device and demonstrate an average singlet-triplet state readout fidelity of 97.59$\%$ by the threshold method. We reveal the inherent deficiency of the threshold method for the relaxed triplet state classification and introduce machine learning as a relaxation-independent readout method to reduce the misjudgment. The readout fidelity for classifying the simulated single-shot traces can be improved to 99.67$\%$ by machine learning method, better than the threshold method of 97.54$\%$ which is consistent with the experimental result. This work indicates that machine learning method can be a strong potential candidate for alleviating the restrictions of stably achieving high-fidelity and high-accuracy singlet-triplet state readout in large-scale quantum computing. 
\end{abstract}

\maketitle


\section{\label{sec:level1}INTRODUCTION}

Electron spin qubit in silicon-based quantum dot (QD) platforms has been a potent way to realize universal quantum computing \cite{zwanenburg2013silicon,Zhang_2018,zhang2019semiconductor}, owing to the long coherence time up to 120 $\upmu$s \cite{veldhorst2014addressable}, high-fidelity single- and two-qubit gates \cite{yoneda2018quantum,xue2022quantum,noiri2022fast,mkadzik2022precision,mills2022two}, and potential for scalability with advanced semiconductor industry \cite{ha2021flexible,zwerver2022qubits}. For large-scale quantum computing, it also requires fast high-fidelity state readout during the qubit coherence time. Compared to the Elzerman readout \cite{elzerman2004single,morello2010single}, Pauli spin blockade (PSB) readout \cite{johnson2005triplet,shaji2008spin,lai2011pauli,maune2012coherent} can perform traditional single-shot readout by monitoring the charge sensor current \cite{maune2012coherent,fogarty2018integrated,jock2018silicon}, but also can achieve gate-based dispersive readout \cite{crippa2019gate,urdampilleta2019gate,west2019gate,zheng2019rapid,noiri2020radio} through an integrated reflectometry circuit to greatly improve the measurement efficiency and minimize the electrode overheads. And benefiting from the excited orbit state \cite{zhao2019single,philips2022universal}, the readout window of PSB is large compared to the reservoir-based Elzerman readout which is limited by Zeeman energy, to achieve high-fidelity state readout.

Fault-tolerant quantum computing requires the state readout fidelity higher than 99$\%$. However, to achieve high-fidelity single-triplet (ST) state readout requires rigorous optimization of the experimental parameters \cite{harvey2018high,connors2020rapid,niegemann2022parity,oakes2023fast}, especially the signal-to-noise ratio (SNR) \cite{blumoff2022fast} and the ratio of the triplet state relaxation timing ($t_{\rm relax}$ which is unique for each single-shot trace of the relaxed triplet state. See Appendix~\ref{App-AA} and Fig.~\ref{fig:11} for more details) to the readout time ($t_{\rm read}$). Keeping these optimizations stable in scalable QD devices to maintain high-fidelity state readout is a challenge, and also the basis for the future implementations of quantum algorithms \cite{watson2018programmable,xue2022quantum} and quantum error correcting protocols \cite{takeda2022quantum,van2022phase}. Currently, the ST state readout fidelity is extracted by the widely used threshold method (THM). THM uses the mean value throughout $t_{\rm read}$ as the judgement \cite{barthel2009rapid} which results in a large error for classifying ST state at small ratio of $t_{\rm relax}$ to $t_{\rm read}$ \cite{harvey2018high,blumoff2022fast}, because the smaller $t_{\rm relax}$ causes the mean value of the relaxed triplet state signal to be closer to the singlet. Therefore, to demonstrate the relaxation-independent state readout method, we introduce machine learning (ML) to alleviate the requirement of $t_{\rm relax}$. Different from THM, ML captures the characteristics of each single-shot trace to classify ST states rather than comparing a mean value with a fixed threshold, enabling more accurate identification of the relaxed triplet state. Its effectiveness for classifying the Elzerman readout traces has been demonstrated in semiconductor QD platforms \cite{matsumoto2021noise,struck2021robust}, whether it is robust on ST state readout needs to be demonstrated. 

In this work, we first tune up the double quantum dot (DQD) device and characterize the PSB region by using modulation measurements based on the lock-in amplifier. Then, we focus on the enhanced latching readout (ELR) mechanism \cite{fogarty2018integrated,harvey2018high,connors2020rapid,oakes2023fast} based on PSB for ST state readout. Compared to the conventional PSB readout by detecting the small charge dipole between two QDs, this method has a considerable signal from the total number of electrons to differ by one \cite{harvey2018high,fogarty2018integrated}. The average ST state readout fidelity of 97.59$\%$ is demonstrated via THM. Furthermore, we illustrate the inherent deficiency of THM based on the simulated single-shot traces and introduce ML to better identify the relaxed triplet state trace. The ST state readout fidelity of THM for classifying the simulated traces is 97.54$\%$ which is consistent with the experiment result, and can be improved to 99.67$\%$ by ML. Supported by both the experimental and simulated results, we propose that ML can be a strong potential candidate for stably achieving high-fidelity and high-accuracy ST state readout in large-scale quantum computing.

\section{\label{sec:level2}DEVICE FABRICATION AND PAULI SPIN BLOCKADE}

\begin{figure}[t]
\includegraphics{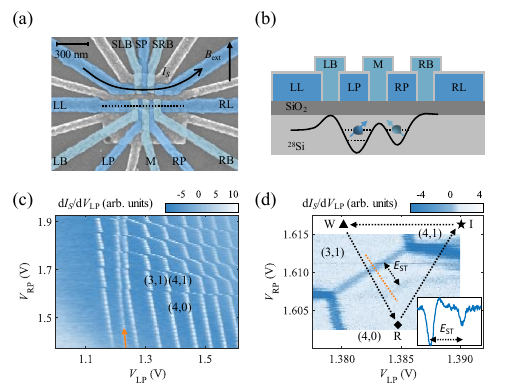}
\caption{\label{fig:1} (a) Scanning electron microscope image of the overlapping gate structure used to define DQD and SET. (b) Cross-section schematic of the device along the dashed line in panel (a). The electrons trapped in two QDs are located under the plunger gates LP and RP. Only one communal reservoir is accumulated under the gate LL. Gates SLB, SRB and SP are used to form SET as the charge sensor and maintain its sensitivity. (c) Charge stability diagram of DQD obtained by sweeping gate voltage $V_{\rm LP}$ and $V_{\rm RP}$. The unexpected charge transition line of the spurious QD (orange arrow) has no effect on the chosen working area around the (3, 1)-(4, 0) interdot charge transition. (d) Three stages of the cyclic pulse sequence where the voltage of stage R is swept to measure the edges of PSB trapezoidal region. Its width is the ST state energy splitting $E_{\rm ST}$ in the (4, 0) configuration. Inset: the average signal of repeated measurement along the orange dotted line to clearly exhibit the triplet state tunnelling event.}
\end{figure}

Fig.~\ref{fig:1}(a) shows the gate layout of the silicon metal-oxide-semiconductor (Si-MOS) DQD device. Overlapping gates \cite{zajac2015reconfigurable,zhang2021controlling,hu2021operation,hu2023flopping} are used to form DQD and a single-electron transistor (SET) as the charge sensor. The tunnel coupling and charge occupation are fine-tuned by the barrier gates (LB, M and RB) and the plunger gates (LP and RP). An in-plane magnetic field of 450 mT is applied to split the three triplet states in energy. Fig.~\ref{fig:1}(b) shows the cross-section schematic of the DQD structure along the dotted line in Fig.~\ref{fig:1}(a). Electrons are trapped in two QDs defined under gates LP and RP (dot L and dot R, respectively). For simplicity, only one communal reservoir under gate LL is accumulated for the electron tunnelling events, and gates RB and RL are grounded. The cross-section schematic of the SET is similar with the DQD, and gates SLB, SRB and SP are used to form a single QD and maintain its sensitivity to detect the electron tunnelling events of the DQD.

Confirmed by the charge detection technique in Ref. \cite{yang2012orbital,hu2021operation}, two QDs can be depleted to the last electron individually with no further charge transition line appearing in the bottom left region, as shown in Fig.~\ref{fig:1}(c). Here, we select the (4, 0) - (3, 1) interdot charge transition as the working area and disregard the unexpected transition line of the spurious QD (orange arrow in Fig. 1(c)), where (\emph{N$_1$}, \emph{N$_2$}) are the number of electrons in dots L and R. We assume that the first two electrons in dot L will occupy the lower energy valley to form a singlet state and do not have any effect on the third and fourth electrons \cite{zhao2019single}. To observe the PSB, we use a cyclic I-W-R pulse sequence (dotted arrows in Fig.~\ref{fig:1}(d)), where the three pulse stages are marked by different symbols and the duration of each stage is variable to measure the relaxation time (see Fig.~\ref{fig:3}(d) and Fig.~\ref{fig:4}(d)). 

First, the device is tuned to the (4, 1) charge configuration at stage I, where the last two electrons in dot L will form an anti-parallel spin pair. Then the DQD is pulsed to the (3, 1) configuration (stage W) to randomly remove one of the four electrons from dot L, either spin down or spin up. Therefore, the spin of the third electron in dot L could form parallel or anti-parallel spin pair with the electron spin in dot R to initialize the ST mixed state. At stage R, we pulse the ST mixed state to (4, 0) to detect the boundary of the PSB region where only the singlet state can tunnel to the (4, 0) configuration. Fig.~\ref{fig:1}(d) plots the edges of the PSB trapezoidal region \cite{maune2012coherent,zhao2019single} by initializing the ST mixed state at stage W and varying the voltage of stage R. Its width is the ST state energy splitting $E_{\rm ST}$ in the (4, 0) configuration. The short edge of the trapezoid is due to the charge transition between the (3, 1) triplet and (4, 0) triplet states. The inset of Fig.~\ref{fig:1}(d) presents the averaged signal of repeated measurement along the orange dotted line to clearly exhibit the triplet state tunnelling event. However, the interdot tunnelling event has no considerable signal to realize a single-shot ST state readout because of the sub-optimal SET position relative to the DQD \cite{fogarty2018integrated}.

\section{\label{sec:level3}ENHANCED LATCHING READOUT}
ELR utilizes the third-party charge configuration which causes the total number of electrons to differ by one, to generate considerable signal than the conventional PSB readout of detecting the charge dipole between two QDs \cite{fogarty2018integrated,harvey2018high}, making single-shot ST state readout possible. The following experiments are based on the single-shot ELR. Under the existing sample conditions, electrons in dot L have a much faster tunnel rate to the communal reservoir than the electrons in dot R under a smaller $V_M$ (see Appendix~\ref{App-A} for more information on the asymmetric coupling conditions). The ST state readout is achieved by monitoring an electron tunnelling event from dot L to the communal reservoir.

\begin{figure}[t]
\includegraphics{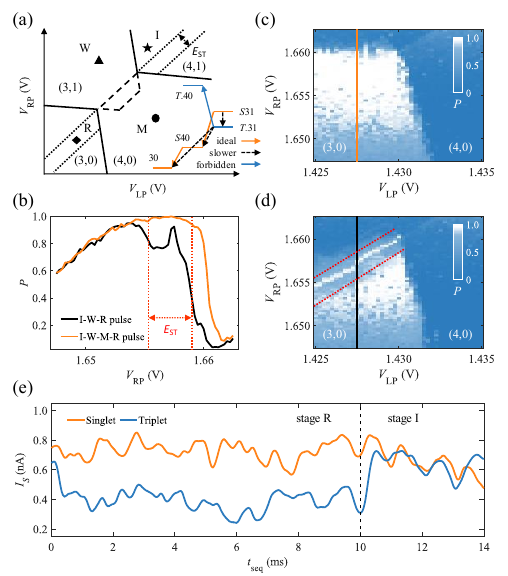}
\caption{\label{fig:2} (a) Schematic charge stability diagram to introduce the enhanced latching region. Inset: energy level and state ladder diagram of the bottom-left enhanced latching region. Only the S(3, 1) state can be mapped to S(4, 0) then to the (3, 0) charge configuration ideally. (b) Readout probability of the tunnelling events as a function of $V_{\rm RP}$ by using I-W-R pulse sequence (black) and I-W-M-R pulse sequence (orange) at $V_{\rm LP}$ = 1.427 V. Two experimental curves are the line cuts of the plot in panel (c) ans (d) at the position of the solid line. Compared to the orange curve, the black curve has a visibly enhanced latching platform due to the forbidden transition of the triplet state. (c-d) Charge stability diagram measured by the two pulse sequences where the voltage of stage R is swept. The measurement starts with an ST mixed state. The orange and black solid lines mark the locations of the experimental results in panel (b), respectively. The red dotted lines show the boundary of the enhanced latching region. (e) Two experimental single-shot traces of the singlet (orange) and triplet (blue) state as a function of the I-W-R pulse sequence time $t_{\rm seq}$. The high-current trace during $t_{\rm seq}$ = 10 ms represents the singlet state, and the low-current trace is the triplet state.}
\end{figure}

Fig.~\ref{fig:2}(a) shows the location of each charge transition line and PSB region around (4, 0) - (3, 1) charge configuration. Two theoretical enhanced latching regions \cite{harvey2018high,zhao2019single} are plotted by the dotted line where the width $E_{\rm ST}$ is the same as the PSB region. We only consider the singlet state $S$ and the ground triplet state $T_-$ for simplification to describe the ELR processes in the theoretical enhanced latching region in (3, 0) configuration. The energy level and state ladder diagram are shown in the inset of Fig.~\ref{fig:2}(a). Here, the orange and blue solid lines represent the ideal and forbidden tunnelling processes, and the black dashed line indicates the tunnel rate in between. Before the readout process of the I-W-R pulse sequence, the initialized state could be (3, 1) singlet or triplet state. When the singlet state is initialized, it will tunnel to the (3, 0) configuration through the S(4, 0) charge state (the ideal process). And a full-one electron signal is generated by the tunnelling event from dot L to the communal reservoir. In contrast, if the initialized state is the triplet state, it cannot tunnel to the (4, 0) charge configuration (the forbidden process) due to the PSB, and the (3, 1) - (3, 0) charge transition is not allowed due to the slower tunnelling process (see Appendix~\ref{App-A}). As a result, no tunnelling event is allowed during stage R. Therefore, singlet and triplet states can be distinguished by monitoring whether an electron can tunnel to the reservoir. Fig.~\ref{fig:2}(e) shows two experimental single-shot traces during the I-W-R pulse sequence time $t_{\rm seq}$ of 14 ms (including stage R and part of stage I), where the high-current level represents the singlet state, and the low-current level is the triplet state.

Next, we measure the readout probability $\emph{P}$ of the tunnelling events from dot L to the communal reservoir to reproduce the theoretical enhanced latching region experimentally. Here, we design an ancillary pulse sequence named I-W-M-R for comparison with the I-W-R pulse sequence. The pulse voltage of stage M (see Fig.~\ref{fig:2}(a)) is deep in the (4, 0) charge configuration; regardless of the kind of initialized (3, 1) charge state, it can ideally tunnel to (4, 0) at stage M and then unobstructed to (3, 0) configuration at stage R. Fig.~\ref{fig:2}(b) plots two typical measurement signals by using I-W-R and I-W-M-R pulse sequences to show the enhanced latching platform, and the corresponding charge stability diagrams are shown in Fig.~\ref{fig:2}(c-d). For the orange curve shown in Fig.~\ref{fig:2}(b), we vary the voltage of stage R of I-W-M-R pulse sequence to measure $\emph{P}$. Based on the unobstructed tunnelling events, $\emph{P}$ rises rapidly to 1 when the voltage of stage R reaches the (3, 0) charge configuration, and the charge stability diagram plotted in Fig.~\ref{fig:2}(a) is effectively reproduced (see Fig.~\ref{fig:2}(c)).However, by using the I-W-R pulse sequence, the black curve in Fig.~\ref{fig:2}(b) shows a clear enhanced latching platform which is due to the forbidden transition of the triplet state in the enhanced latching region. Due to the initialization of the mixed ST state, $\emph{P}$ around the enhanced latching platform should theoretically be 0.25. However, $\emph{P}$ is approximately 0.7 in Fig.~\ref{fig:2}(b), which is mainly caused by the relaxation of the triplet state during the readout time $t_{\rm read}$. In Fig.~\ref{fig:2}(d), different from the charge stability diagram shown in Fig.~\ref{fig:2}(c), the charge transition line of dot R disappears due to the slower tunnelling process from the (3, 1) to (3, 0) charge configuration, which is replaced by a boundary parallel to the interdot transition line. The red dotted lines in Fig.\ref{fig:2}(d) show the boundary of the enhanced latching region. The readout probabilities decrease along the reversed direction of the $V_{\rm RP}$ because the SET current is affected by $V_{\rm RP}$ \cite{yang2011dynamically} as well as the suitable threshold due to capacitive coupling.


\section{\label{sec:level4}READOUT FIDELITY VIA THM}

\begin{figure}[b]
\includegraphics{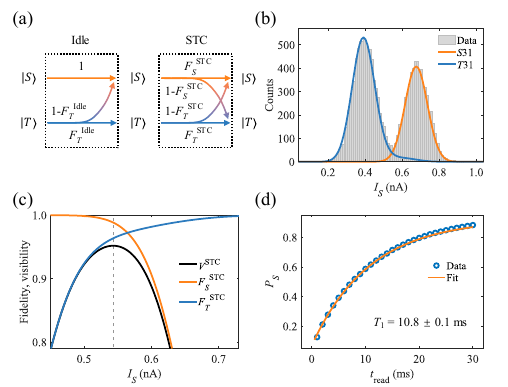}
\caption{\label{fig:3} (a) Evolution of the qubit states during stage R, consisting of two parts: Idle and STC. (b) Histogram of the single-shot measurements integrated over $t_{\rm read}$ = 1.18 ms, with Gaussian fits indicating singlet (orange) and triplet (blue) readout signals. (c) Visibility and individual readout fidelities for singlet and triplet states as a function of $I_S$. The maximum visibility is 95.18$\%$, and the corresponding readout fidelity for the singlet (triplet) state is 98.81$\%$ (96.37$\%$). (d) The measured state relaxation time of approximately 10.8 $\pm$ 0.1 ms by varying $t_{\rm read}$.}
\end{figure}

With the ability to classify the ST states, we next investigate the readout fidelity by using the I-W-R pulse sequence. The evolution of ST states during stage R can be divided into two parts: Idle and state-to-charge conversion (STC), as presented in Fig.~\ref{fig:3}(a). The idle time is caused by the measurement circuit response (see Fig.~\ref{fig:6}(a) and Appendix~\ref{App-B} for details), which limits the experimental data for the fidelity analysis to start at 0.5 ms. The triplet state relaxation during the idle time will cause an readout error, and this error is dependent on the relaxation time $T_1$. Here, $T_1$ = 10.8 $\pm$ 0.1 ms is extracted in Fig.~\ref{fig:3}(d) by using the I-W-R cyclic pulse sequence and varying $t_{\rm read}$ during stage R as the only independent variable \cite{zhang2020giant} to measure the singlet state probability. Therefore, the idle time results in an error of approximately 2.26$\%$ \cite{blumoff2022fast}. For the STC, parts of the triplet state traces will relax to the singlet states, and the remaining triplet state and the singlet state may also be misjudged due to the low SNR. 

Fig.~\ref{fig:3}(b) plots the distribution of the mean values of each readout trace during $t_{\rm read}$ = 1.18 ms (the time range from 0.5 ms to 1.18 ms due to the idle time). The high-current signal represents the singlet state, and the low-current signal is the triplet state. To calculate the readout fidelity, we fit the bimodal distribution shown in Fig.~\ref{fig:3}(b) based on two noise-broadened Gaussian distributions with an additional relaxation term from the triplet state during $t_{\rm read}$ \cite{barthel2009rapid}. We repeat the fidelity analysis for various integration times and select an optimal threshold SET current $I_S$ from the fitted parameters to calculate the optimal readout fidelity. The maximum STC visibility $V^{\rm STC}$ = 95.18$\%$ is shown in Fig.~\ref{fig:3}(c) with an readout time $t_{\rm read}$ = 1.18 ms and an optimal $I_S$ of approximately 0.54 nA. The corresponding readout fidelity for the singlet (triplet) state is 98.81$\%$ (96.37$\%$) with an average STC fidelity of $F^{\rm STC}$ = 97.59$\%$, which is mainly limited by the triplet state relaxation and the small SNR (due to the fluctuation of $I_S$, see Fig.~\ref{fig:6}(b) and Appendix~\ref{App-B}). The STC fidelity for various $t_{\rm read}$ is shown in Fig.~\ref{fig:6}(d). To further extrapolate our results by shortening $t_{\rm read}$ and enhancing the SNR, the $F^{\rm STC}$ can be improved to more than 99$\%$ or even 99.9$\%$ (see Appendix~\ref{App-B} for more details).

\section{\label{sec:level5}READOUT FIDELITY VIA ML}
\begin{figure}[b]
\includegraphics{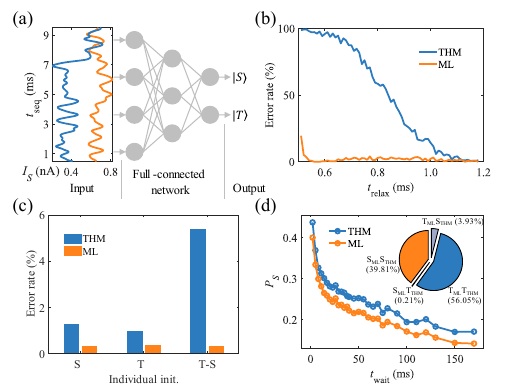}
\caption{\label{fig:4} (a) The classification procedure consisting of three fully connected networks. The output is two categorical indices, marking either the singlet or triplet state. (b) Error rates of judging the relaxed triplet state under a series of $t_{\rm relax}$ at $t_{\rm read}$ = 1.18 ms. (c) Error rates of ML (orange) and THM (blue) for classifying different initialization states, where the chosen $t_{\rm read}$ is consistent with (b). (d) Readout probability of the singlet state classified by ML (orange) and THM (blue). Inset: classification results of the ST states at $t_{\rm wait}$ = 2 ms, where the left label of notation represents the result judged by ML and the right label is from THM.}
\end{figure}

Due to the inherent deficiency of THM, i.e., using a mean value throughout $t_{\rm read}$ to determine whether it exceeds a fixed threshold, tunnelling events with a smaller $t_{\rm relax}$ relative to $t_{\rm read}$ will lead to the misjudgment of the triplet state as the singlet. Because the smaller $t_{\rm relax}$ results in more high-current signals during one single-shot trace, which causes the mean value of the relaxed triplet state during $t_{\rm read}$ to be closer to the singlet state signal (see Appendix~\ref{App-D}). Here, we introduce ML \cite{matsumoto2021noise,struck2021robust} to classify the relaxed triplet state more accurately. The ML network shown in Fig.~\ref{fig:4}(a) consists of three full-connected linear networks, which is sufficient to classify the ST signals because the experimental single-shot traces have only three possible situations during $t_{\rm read}$: remaining at the high-current level (the singlet state) or the low-current level (the triplet state without relaxation), or switching from the low- to the high-current level (the relaxed triplet state). The first two fully-connected linear networks are followed by the sigmoid activation function, and the softmax function is added at the end to output the classification results. 

For ML network training, we use Monte Carlo method to simulate single-shot trace with Gaussian noise and SET fluctuation noise equivalent to the experimental SNR (see Fig.\ref{fig:7}(c)) and $F^{\rm STC}$ (see Fig.\ref{fig:8}). Appendix~\ref{App-C} describes more information on the generation and the quality of simulated traces. The training set includes 480000 simulated traces and the validation set includes 90000 traces. We choose the simulated trace during $t_{\rm read}$ = 9.5 ms (the time range from 0.5 ms to 9.5 ms) as the input, and the output is two categorical indices, marking either the singlet or triplet state.

To highlight the advantage of ML in classifying the relaxed triplet state, we regenerate one million sets of simulated traces (the testing set for ML) and count the error rates. Here, the error rate is defined as the fraction of times when the classification result is different from the preassigned label. And the integrated time of $t_{\rm read}$ = 1.18 ms is chosen to be consistent with the fidelity analysis range shown in Fig.~\ref{fig:3}(b). In Fig.~\ref{fig:4}(b), the error rate of THM for classifying the relaxed triplet state increases with decreasing $t_{\rm relax}$ due to its data processing scheme. In contrast, the error rate of ML is very low in a wide $t_{\rm relax}$ range and is nearly independent of $t_{\rm relax}$ which indicates a relaxation-independent readout method (see Appendix~\ref{App-E} for the comparison with $t_{\rm read}$ = 1.5 ms and 2.0 ms to further validate the deduction). The error rate of ML rises rapidly at a small $t_{\rm relax}$, which may be due to the insignificant rising edge of the signal under the influence of noise. We speculate that if the sampling rate (here, 100 kSa/s) is further increased to provide more information on the rising edge, the error rate of ML will remain low during the entire range of $t_{\rm relax}$. Fig.~\ref{fig:4}(c) shows the error rates of THM and ML for classifying the different initialized states. The symbols ‘S’, ‘T’ and ‘T-S’ represent the simulated traces for the singlet state, the triplet state without relaxation and the relaxed triplet state, respectively. The error rates of ML are much lower than that of THM for all initialized states, especially for the relaxed triplet state. Here, the readout fidelity of the THM is approximately 97.54$\%$ which is consistent with the experimental result of 97.59$\%$, and can be improved to 99.67$\%$ by using ML to reidentify the misclassified state. Furture work can use more sophisticated neural network architecture including convolutional neural network \cite{szegedy2015going} and the long short-term memory \cite{schmidhuber1997long} to achieve higher classification accuracy.

Finally, we reprocess the experimental traces by using ML to extract the relaxation time in the (3, 1) configuration. Fig.~\ref{fig:4}(d) shows the discrimination results of the singlet state probability $P_S$ as a function of the wait time $t_{\rm wait}$ at stage W. Compared with THM, $P_S$ judged by ML is generally reduced, which is mainly from the reidentification of the relaxed triplet state. The inset of Fig.~\ref{fig:4}(d) shows the classification of these two methods at $t_{\rm wait}$ = 2 ms, where the left label of notation represents the result judged by ML and the right label is from THM. Compared to THM, approximately 3.93$\%$ of the experimental traces is reidentified from the misclassified singlet state to the triplet, reflecting the advantages of ML and proving its superiority in achieving the high-accuracy ST state readout.

\section{\label{sec:level6}DISCUSSION AND CONCLUSIONS}
For ST state readout fidelity, there are two key points about THM: $t_{\rm relax}/t_{\rm read}$ and SNR. The large $t_{\rm relax}$ and short $t_{\rm read}$ are benefit to the readout fidelity of THM for classifying the relaxed triplet state. And the SNR should be considered together to improve the readout fidelity of the singlet state and the triplet state without relaxation. Although the rigorous optimization for these factors has been made experimentally on a few qubits, it is a challenge to ensure that all qubits are stable and well-optimized for large-scale quantum computing. Therefore, ML is introduced to alleviate the requirement of $t_{\rm relax}$, and its relaxation-independent property has been demonstrated in Section ~\ref{sec:level5}. Here, ML can not only classify the relaxed triplet state with high-accuracy (relaxation-independent property), but also effectively classify the singlet state and the triplet state without relaxation (see Fig.~\ref{fig:4}(c)). It indicates that ML is also robust to noise during single-shot trace, however, its noise-resilient needs to be validated in the future. Therefore, ML could be an robust candidate for high-fidelity and high-accuracy ST state readout without the need for well-optimized experimental parameters, which reduces the difficulty of achieving fault-tolerant quantum computing.

It is worth emphasizing that ELR has some significant advantages over Elzerman readout. Firstly, there are only two tunnel rates during the ELR process, a fast one and a slow one, which can be easily achieved in gate-defined QD device. But for Elzerman readout, we must tune the tunnel out rates of each spin state and the reload rate of spin down electron. And the relationship between these tunnel rates needs to be optimized carefully. Then, due to the lack of electron reload process, the ELR mechanism does not take into account the effect of dark count. Thirdly, the filter cutoff frequency only needs to balance $t_{\rm read}$ and idle time, but the Elzerman readout also needs to consider the reload rate to allow the fast change in electron occupancy to be detectable. Further, benefiting from the much larger valley splitting or orbit energy difference compared to the Zeeman splitting, ELR mechanism works well under high electron temperature \cite{yang2020operation,petit2020universal,niegemann2022parity,oakes2023fast}. It also works well under low external magnetic field \cite{zhao2019single} to simplify the requirement of microwave engineering \cite{scarlino2015second}. And ELR mechanism not only can distinguish the ST states, but also can achieve single electron spin state readout if there is a reference spin \cite{zheng2019rapid} or implementation of two-qubit gates \cite{leon2021bell,philips2022universal}.

In summary, we experimentally classify the singlet-triplet state by the enhanced latching readout mechanism based on Pauli spin blockade in a silicon double quantum dot device. A singlet-triplet average readout fidelity of 97.59$\%$ is calculated by the widely used threshold method, and the poor fidelity is from the low signal-to-noise ratio and small ratio of the relaxation time to the readout time. Further, we have revealed the inherent deficiency of the threshold method for the misjudgment of the relaxed triplet state, and introduce machine learning as a relaxation-independent readout method to reidentify this state. We achieve a readout fidelity of 99.67$\%$ by using machine learning for classifying the simulated traces compared to the threshold method of 97.54$\%$, and this fidelity could be further increased with a more sophisticated neural network architecture. Our study indicates that machine learning can be a strong potential candidate for alleviating the restrictions of stably achieving high-fidelity and high-accuracy singlet-triplet state readout in large-scale quantum computing.

\begin{acknowledgments}
This work was supported by the National Natural Science Foundation of China (Grants No. 12074368, 92165207, 12034018 and 92265113), the Innovation Program for Quantum Science and Technology (Grant No. 2021ZD0302300), the Anhui Province Natural Science Foundation (Grants No. 2108085J03), and the USTC Tang Scholarship. This work was partially carried out at the USTC Center for Micro and Nanoscale Research and Fabrication.
\end{acknowledgments}

\appendix

\section{\label{App-AA}The relaxation timing}
\begin{figure}[h]
\includegraphics{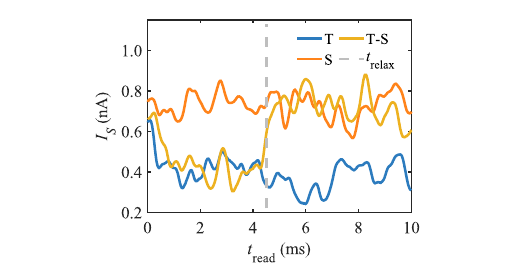}
\caption{\label{fig:11} Three experimental traces of the singlet state (‘S’), the triplet state without relaxation (‘T’) and the relaxed triplet state (‘T-S’) during stage R. $t_{\rm relax}$ is the relaxation timing of individual trace of the relaxed triplet state.}
\end{figure}

Fig.~\ref{fig:11} plots three typical single-shot traces of the singlet state (‘S’), the triplet state (‘T’) and the relaxed triplet state (‘T-S’). The traces of the singlet and triplet states are the same as the one shown in Fig.~\ref{fig:2}(e). Here the relaxation timing $t_{\rm relax}$ marks the location that the readout trace switches from the low-current level to the high-current level, which is unique to the relaxed triplet state. And $t_{\rm relax}$ is unique for every single-shot trace of the relaxed triplet state, while the relaxation time is extracted from the statistical distribution of all $t_{\rm relax}$.

\begin{figure}[b]
\includegraphics{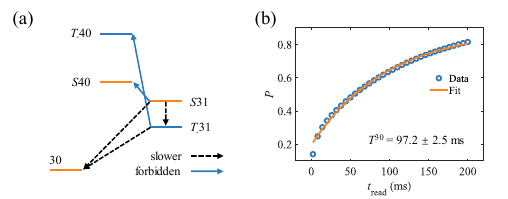}
\caption{\label{fig:5} (a) The energy level and state ladder diagram at the right side of the enhancement platform with $V_{\rm RP}$ = 1.66 V in Fig.~\ref{fig:2}(b), where no tunnelling event occurs except the relaxation between (3, 1) and (3, 0) charge configuration. (b) The relaxation time between the (3, 1) and (3, 0) states to confirm the asymmetric coupling condition. The relaxation time of approximately 97.2 $\pm$ 2.5 ms is extracted.}
\end{figure}

\section{\label{App-A}The asymmetric coupling condition}
We measured the relaxation time $T^{30}$ between the (3, 1) and (3, 0) states to confirm the asymmetric coupling condition. We use the I-W-R pulse sequence and set the voltage of stage R to the right side of the enhancement platform ($V_{\rm RP}$ = 1.66 V in Fig.~\ref{fig:2}(b)). Figure~\ref{fig:5}(a) presents the energy level and state ladder diagram at stage R. The energy of S(4, 0) state is higher than any (3, 1) states, therefore, the initialized (3, 1) mixed state can not tunnel to (4, 0) state (the forbidden process). And due to the much slower tunnel rate between dot R and the communal reservoir, the (3, 1) mixed states can not tunnel to (3, 0) charge configuration directly. Figure~\ref{fig:5}(b) shows the readout probability of the tunnelling event from (3, 1) to (3, 0) configuration with increasing $t_{\rm read}$, and the fitted relaxation time is approximately 97.2 $\pm$ 2.5 ms, which has little effect on the readout process.

On the contrary, during the ELR experiment, the electron transfer is done at the beginning of stage R on the order of hundreds of kHz to MHz or even faster due to the ideal tunnelling process between (3, 1) and (4, 0) charge state (see the state ladder diagram shown in Fig.~\ref{fig:2}(a)), which is unmeasurable due to the readout bandwidth.

\begin{figure}[b]
\includegraphics{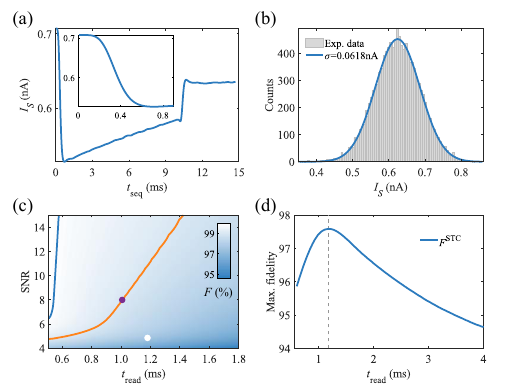}
\caption{\label{fig:6} (a) Averaged readout signal during stage R and part of stage I. The falling edge is clearly shown in the inset. (b) Histogram of the single-shot measurements during 12.00 ms – 12.68 ms in (a) with a variance of 0.0618 nA, which is comparable to the one shown in Fig. 3(b) in the main text with a variance of approximately 0.0621 nA. (c) Extrapolated readout fidelity at different $t_{\rm read}$ and SNR. The blue and orange lines mark where the values of $F^{\rm STC}$ are 99$\%$ and 99.9$\%$, respectively. The white circle indicates the experimental condition shown in Fig. 3(b) in the main text, and the purple circle represents the readout fidelity of 99$\%$ with $t_{\rm read}$ = 1.0 ms and SNR = 8. (d) The STC fidelity for various readout times.}
\end{figure}

\section{\label{App-B}Optimization of the readout fidelity analysis}
One limitation of our measurement is caused by the idle time before the fidelity analysis compared to the triplet state relaxation time $T_1$. The falling edge lasting $\sim$ 0.5 ms is clearly shown in the inset of Fig.~\ref{fig:6}(a), resulting in an error of approximately 2.26$\%$ due to the triplet state relaxation. It can be optimized by using RF coaxial cables to achieve ST state readout which is benefiting to the readout bandwidth. Radio-frequency reflectometry is also a popular way to decrease $t_{\rm read}$ to alleviate the effect of the short $T_1$. 

On the other hand, due to THM's characteristic of using the mean value to classify the ST state, the fluctuation of the SET current between two individual experiment trace greatly limits the readout fidelity by lowering the SNR of the bimodal distribution. Figure~\ref{fig:6}(b) shows the histogram of the average single-shot measurement signals during 12.00 ms – 12.68 ms (stage I) in Fig.~\ref{fig:6}(a), with a variance of 0.0618 nA, which is comparable to the variance in Fig.~\ref{fig:3}(c) of approximately 0.0621 nA. Therefore, keeping $I_S$ stable or finding a way to eliminate its effect will significantly improve the readout fidelity of THM.

To extrapolate our results, Fig.~\ref{fig:6}(c) shows the calculated readout fidelity $F^{\rm STC}$ under different SNR and $t_{\rm read}$ by using part of the fitted parameters in Fig.~\ref{fig:3}(b). The orange line represents a set of SNR and $t_{\rm read}$ that achieving $F^{\rm STC}$ = 99$\%$, and the blue line is for $F^{\rm STC}$ = 99.9$\%$. The white circle indicates the experimental condition shown in Fig.~\ref{fig:3}(b): $t_{\rm read}$ = 1.18 ms and SNR = 4.52. When $t_{\rm read}$ is decreased to 1.0 ms and the SNR is increased to 8, the readout fidelity can reach 99$\%$, which is marked by the purple circle in Fig.~\ref{fig:6}(c).

\section{\label{App-C}Experiment data simulation}

\begin{figure}[t]
\includegraphics{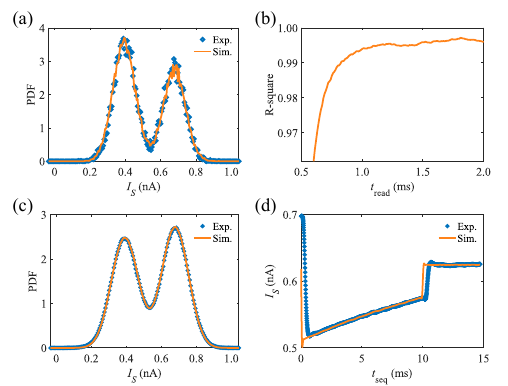}
\caption{\label{fig:7} A Monte Carlo simulation of 100, 000 single-shot traces with added Gaussian noise and SET fluctuation noise to reproduce the experimental signal. (a) The distributions of the mean data during $t_{\rm read}$ = 1.18 ms. (b) The simulation quality of the mean data as a function of $t_{\rm read}$. (c) The distribution of the all data between 0.5 ms and 9.5 ms. (d) The averaged single-shot trace during stage R and part of stage I.}
\end{figure}

To simulate the distribution of the experimental data, we first fit the distribution of all data during $t_{\rm read}$ = 9.5 ms with the Gaussian Mixture Model (GMM) to obtain the mean and variance of two signal peaks. Then combining the experimental parameters and using the Monte Carlo method, the simulated signals with added Gaussian noise and SET fluctuation noise are generated to reproduce the experimental signals. The simulated traces are divided into three types: the singlet state, the triplet state without relaxation and the relaxed triplet state. Here the SET fluctuation noise is extracted from Fig.~\ref{fig:6}(b) to reproduce the SET current fluctuation between two individual experiment trace. Fig.~\ref{fig:7}(a) shows the distributions of the mean value during $t_{\rm read}$ = 1.18 ms where the experimental data shown in Fig.~\ref{fig:3}(b) (the blue circles) is in good agreement with the simulated results (the orange line). The simulation quality with a high value of R-square is shown in Fig.~\ref{fig:7}(b). Here, the R-square is low at small $t_{\rm read}$ which is due to signal fluctuation during a short integrated time.The distribution of the all data during $t_{\rm read}$ = 9.5 ms is shown in Fig.~\ref{fig:7}(c) and Fig.~\ref{fig:7}(d) is the comparison of the averaged readout signal during stage R where the falling edge is not well reproduced due to the lack of information on the circuit response. All of these simulated results are in good agreement with the experiment.

\begin{figure}[h]
\includegraphics{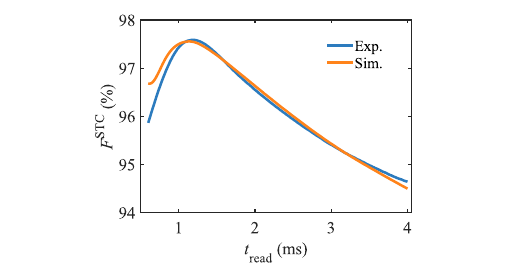}
\caption{\label{fig:8} The readout fidelity $F^{\rm STC}$ of the experimental and the simulated traces for various $t_{\rm read}$. The values and trends of the simulated traces is in good agreement with the experimental results, which indicates the quality of the simulated traces.}
\end{figure}

Figure~\ref{fig:8} plots the comparison of the readout fidelity $F^{\rm STC}$ between the experimental results (the blue curve, also see Fig.~\ref{fig:6}(d)) and the simulated results (the orange curve) calculated by THM. Except for the short $t_{\rm read}$ where the simulation quality of the R-square shown in Fig.~\ref{fig:8}(b) is low, the values and trends of the readout fidelity for classifying the simulated traces is in good agreement with the experimental results, which indicates the quality of the simulated traces.

\section{\label{App-D}The mean value through $t_{\rm read}$}
\begin{figure}[h]
\includegraphics{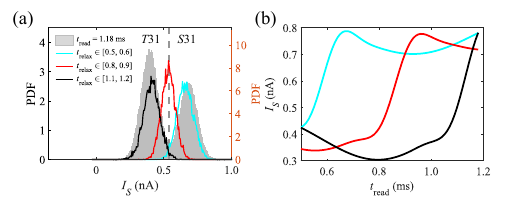}
\caption{\label{fig:9} (a) The distributions of the mean data during $t_{\rm read}$ = 1.18 ms. The bimodal distribution is the same as the simulated results shown in Fig.~\ref{fig:7}(a). The other three distributions are from the relaxed triplet state with different $t_{\rm relax}$. The bright blue is close to the distribution of the singlet state, therefore, its error rate is high. (b) Three simulated traces of the relaxed triplet state with different $t_{\rm relax}$. The color of each curve corresponds to the one in (a).}
\end{figure}

Here, we explain in detail the calculation of the mean value through $t_{\rm read}$, and why THM results in large error for classifying the relaxed triplet state at the small ratio of $t_{\rm relax}$ to $t_{\rm read}$. Figure~\ref{fig:9}(a) shows the distributions of the mean data during $t_{\rm read}$ = 1.18 ms. The gray signal is the statistical bimodal distribution of the mean values of all simulated traces. The other three traces are the distributions of the relaxed triplet state signals with different $t_{\rm relax}$ in order to illustrate the relationship between the error rate and $t_{\rm relax}$. Here, most of the bright blue signals are located to the right of the threshold, which will be misclassified by THM. On the contrary, due to the larger $t_{\rm relax}$ than the bright blue and red signals, the black signal has the lowest error rate. Figure~\ref{fig:9}(b) plots three sample single-shot traces of the simulated triplet state with different $t_{\rm relax}$ selected from Fig.~\ref{fig:9}(a). The color of each curve corresponds to that in Fig.~\ref{fig:9}(a). For a single-shot trace, the mean value of all data points during $t_{\rm read}$ = 1.18 ms is used to compare with a fixed threshold to determine whether it is the singlet state or the triplet state. Here, for the bright blue trace, the ratio of $t_{\rm relax}$ $\in$ (0.5 ms, 0.6 ms) to $t_{\rm read}$ = 1.18 ms is smaller than the red ($t_{\rm relax} \in$ (0.8 ms, 0.9 ms)) and black ($t_{\rm relax} \in$ (1.1 ms, 1.2 ms)) traces. Therefore, its mean value is close to the right peak of the bimodal distribution, i.e., the singlet state signal, which results in a large error rate.

\section{\label{App-E}The error rate for classifying the relaxed triplet state}
\begin{figure}[h]
\includegraphics{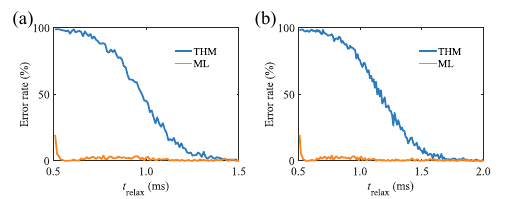}
\caption{\label{fig:10} The error rate of judging the relaxed triplet state with $t_{\rm read}$ = 1.5 ms (a) and 2.0 ms (b). The error rate of the THM increases with decreasing relaxation time $t_{\rm relax}$, and for ML, it’s nearly independent of $t_{\rm relax}$.}
\end{figure}

Figure~\ref{fig:10} presents the error rate for classifying the relaxed triplet state with $t_{\rm read}$ = 1.5 ms and 2.0 ms. Since ML uses all data points of the trace to classify the state, the error rate is nearly independent at any $t_{\rm relax}$. However, the THM uses the mean value throughout $t_{\rm read}$ as the judgment signal, which will suffer from large errors for classifying the relaxed triplet state at small ratios of $t_{\rm relax}$ to $t_{\rm read}$. Therefore, the larger $t_{\rm read}$ is, the more signals of the relaxed triplet state will be misjudged due to the small $t_{\rm relax}$.

\bibliographystyle{unsrt}
\bibliography{apssamp}

\end{document}